\documentclass[10pt, conference, compsocconf]{IEEEtran}
\usepackage{graphicx}
\usepackage{listings}

\newcommand{\specialcell}[2][c]{%
	\begin{tabular}[#1]{@{}c@{}}#2\end{tabular}}
\begin{document}
\title{Porting the microphysics model CASIM to GPU and KNL Cray machines}
	
\author{\IEEEauthorblockN{Nick Brown, Alexandr Nigay, Michele Weiland}
	\IEEEauthorblockA{EPCC, The University of Edinburgh, \\ James Clerk Maxwell Building, \\ Peter Guthrie Tait Road, Edinburgh}	
	\and
	\IEEEauthorblockN{Adrian Hill, Ben Shipway}
	\IEEEauthorblockA{UK Met Office, FitzRoy Road, Exeter, Devon}
		
}
\maketitle
	
\begin{abstract}
CASIM is a microphysics scheme which calculates the interaction between moisture droplets in the atmosphere and forms a critical part of weather and climate modelling codes. However the calculations involved are computationally intensive and so investigating whether CASIM can take advantage of novel hardware architectures and the likely increase in performance this might afford makes sense.

In this paper we present work done in porting CASIM to GPUs via the directive driven OpenACC and also modifying CASIM to take advantage of the Knights Landing (KNL) processor using OpenMP. Due to the design, models extracting out specific computational kernels for offload to the GPU proved suboptimal and instead the entire scheme was ported over to the GPU. We consider the suitability and maturity of OpenACC for this approach as well as important optimisations that were identified. Enabling CASIM to take advantage of the KNL was significantly easier, but still required careful experimentation to understand the best design and configuration. The performance of both versions of CASIM, in comparison to the latest generation of CPUs is discussed, before identifying lessons learnt about the suitability of CASIM and other similar models for these architectures. The result of this work are versions of CASIM which show promising performance benefits when utilising both GPUs and KNLs and enable the communities to take advantage of these technologies, in addition to  general techniques that can be applied to other similar weather and climate models.
\end{abstract}
	
	
\section{Introduction}
The Cloud AeroSol Interactions Microphysics (CASIM) model \cite{kid} is a bulk microphysics scheme, calculating interactions between moisture droplets in the atmosphere. Modelled at the millimetre scale, these droplets represent moisture in many different states such as vapour, liquid water, snow, ice and graupel. The modelling of moisture is a crucial aspect of weather and climate codes, and this scheme is designed to be used as a sub-model by other more general parent models. Current parent models that utilise CASIM are the Met Office Unified Model (UM), Met Office NERC Cloud Model (MONC) \cite{easc}, Large Eddy Model (LEM) \cite{lem} and Kinematic Driver (KiD) \cite{kid} model.The calculations performed by CASIM are computationally intensive and, on homogeneous machines, such as ARCHER (an XC30) or MONSooN (an XC40), a significant portion of the overall model runtime is taken up by CASIM which in some cases can double or even triple the entire runtime \cite{slowcasim}.

Scientists are driving computational models harder and harder, requiring runs at greater resolution and/or near real time to achieve their science. This requires large amounts of computational resource and because CASIM is in use by many different parent models an important question to answer is whether the use of novel hardware architectures, such as GPUs or Knights Landing (KNL), can improve performance. OpenACC has been used as the technology to support CASIM on the GPU, but because of the make up of the code the decision was made to offload the entirety of the scheme including all computation, conditionals and loops onto the GPU which involved porting over 123 Fortran subroutines across 50 modules. This approach was adopted to minimise the cost of data transfer and enable the CPU to concurrently work on other parts of the atmospheric model whilst CASIM is being executed by the GPU. 

Because the KNL supports both MPI and OpenMP codes it is an attractive target for many models. Whilst the performance benefits of previous generations of this technology have been varied, the latest generation are a significant improvement over Knights Corner and many people have reported positive results. The accessibility of programming the KNL makes it a natural target for CASIM and we have used OpenMP to add threading in order to fully take advantage of this hardware.

A key question to answer has been where CASIM runs best, and hence where the focus for future technology should be. Is it advantageous to utilise Cray machines with accelerators such as GPUs or many core CPUs such as the KNL? Alternatively is a strategy based upon utilising the latest generation of CPUs optimal? Section 2 lays the foundation for this work, it introduces both the CASIM microphysics scheme in more detail and discusses previous GPU acceleration work that we have performed on one of the parent models, MONC. Section 3 focusses on the GPU version of CASIM, we discuss the porting of the scheme using OpenACC, performance on Piz Daint both in its current XC50 (with P100s) configuration and previous generation XC30 (K20X) configuration, before going on to discuss optimisations and technical challenges encountered along with their workarounds. Section 4 discusses the KNL version of CASIM, providing a description of the work done to optimise CASIM for this technology and then performance comparisons on the ARCHER KNL system (an XC40.) In section 5 we contrast and compare the GPU and KNL results against CPU only runs of CASIM on both Haswell (XC50) and Broadwell (XC40) before drawing some conclusions and discussing future work in section 6.

\subsection{Machine configurations}
In this paper a variety of different machines and technologies are used for evaluation, the specifications of these are summarised in this section.
\begin{itemize}
\item \textbf{Piz Daint (Cray XC50):} each node contains one Intel Xeon E5-2690 version 3 CPU (Haswell), a NVIDIA Tesla P100 GPU and 64GB RAM. Each processor contains 12 physical cores, each at base clock frequency of 2.6 Ghz with 64KB level one cache and 256KB level two cache. There is 30MB shared level three cache on the package. The P100 GPUs fitted to Piz Daint also contain 16GB of on-chip CoWoS HBM2 Stacked Memory.
\item \textbf{Piz Daint (\textit{until September 2016}) (Cray XC30):} each node contained one Intel Xeon E5-2670 CPU (SandyBridge), a NVIDIA Tesla K20X and 32GB RAM. Each processor contained 8 physical cores, each at base clock frequency 2.6 Ghz with 64KB level one cache and 256KB level two cache. There was 20MB shared level three cache on the package. The K20X GPUs also contained 6GB of on-chip GDDR5 memory.
\item \textbf{MONSooN (Cray XC40):} each node contains two Intel Xeon E5-2695 version 4 CPUs (Broadwell) and 64GB RAM. Each processor contains 18 physical cores, each at base clock frequency of 2.1 Ghz with 64KB level one cache and 256KB level two cache. There is 45MB shared level three cache on the package.
\item \textbf{ARCHER (Cray XC30):} each node contains two Intel Xeon E5-2697 version 2 (Ivy Bridge) CPUs and 64GB RAM. Each processor contains 12 physical cores, each at base clock frequency of 2.7 Ghz with 64KB level one cache and 256KB level two cache. There is 30MB shared level three cache on the package.
\item \textbf{ARCHER KNL (Cray XC40):} each node contains one 7210 Knights Landing (KNL) CPU with 64 cores and 96GB RAM. Each core has a clock frequency of 1.30 Ghz, 64KB of level one cache and with two cores sharing a 1MB level two cache. There is an additional 16GB MCDRAM on-chip which can be used as either a level three cache or as direct main memory from code.
\end{itemize}

\section{Background}
Microphysics is traditionally very computationally intensive and as such certain design decisions are made to trade off the accuracy of solution verses the computational resources required. Broadly these models can be split into \emph{bulk} and \emph{bin} schemes. In bulk schemes all processes are defined as specific parameters where a function defines the distribution of particles and four key parameters for each particle: it's diameter (D), the particle's shape ($\mu$), the particle's slope ($\lambda$) and distribution inception on the Y axis (No.) Whilst all bulk models use these parameters and the same equations, they change the computational complexity by fixing some and varying others, this is known as the moment. For instance in a single moment scheme a particle's diameter (D) is defined and all other variables are fixed, in a double moment scheme both D and No are varied and the other two parameters fixed. In a triple moment scheme D, No and $\mu$ are varied and the slope fixed. In contrast bin schemes make fewer assumptions as they explicitly solve equations that govern the size of droplets by concentrating on drop size distributions. Based on their explicit size particles are discretised into a number of bins, hence the name of the approach, which the equations are solved upon. Bin and bulk schemes have different pros and cons, both in terms of computational requirements and also situations where they produce the most accurate results. 

CASIM is a three moment bulk scheme, most importantly it handles aerosol in clouds in an entirely innovative way which provides far more accuracy with these schemes than previous generations. The role of aerosol is one of the major uncertainties of the hydrological process and thermal radiation. CASIM has explicit activation schemes that transform aerosol to clouds, but very importantly once activated it will carry this aerosol mass in the cloud which allows one to study the evaporation of aerosol and how moisture is returned to the system. This important advancement means that CASIM provides more accurate results in many cases, but the cost is that it is more computationally intensive than peer 3 moment bulk models. The parent model, which drives CASIM, will decompose three dimensional space into columns and CASIM works on each column independently as all processes being simulated are limited to the vertical. This therefore means that the microphysics modelling of one column, whilst it is tightly coupled in the vertical, is entirely independent of other columns in the other two dimensions. Proceeding in sub-timesteps of the parent model, CASIM performs a number of computationally intensive algorithms such as accretion, which calculates how droplets combine together and accelerate as they fall, auto conversion which maps moisture between states (such as from cloud to rain) and sedimentation which determines whether vapour droplets are large enough to fall through the atmosphere. As an input the parent model will pass CASIM many different variables, including 3D prognostic (raw data) fields representing moisture in different states. Broadly speaking there are two major configurations to the scheme running warm, where one only considers moisture above freezing point and also running cold which models moisture in all states both above and below freezing point. Parent models maintain a number of \emph{q} 3D prognostic fields, which represent moisture at each point on the grid in different states. These are worked on directly by the scheme and the warm configuration involves 5 \emph{q} fields, whereas the cold configuration involves 18 \emph{q} fields. Hence the cold configuration involves far more computation.

\subsection{MONC}
\label{sec:moncbg}
The Met Office NERC Cloud model (MONC) is a newly developed open source, high resolution modelling framework. Employing large eddy simulation, this model studies the physics of turbulent flows and enables scientists to further develop and test physical parametrisations and assumptions used in numerical weather and climate prediction. The standard homogeneous version of this code is designed to be run on many thousands of cores and has demonstrated good performance and scalability on up to 32768 cores \cite{easc} on XC30 and XC40 Cray machines. MONC has been designed around pluggable components where the majority of the code complexity, including all of the science and parallelisation, are contained within independent units. Splitting the model into these units, for instance a component for advection (calculating movement due to wind), another for calculating buoyancy terms and another for calculating pressure means that it is very easy both to configure the model by plugging in some and unplugging other components, and also develop new functionality independent of other parts of the model. It is also possible to call out to other models for specific aspects (such as CASIM for microphysics) by developing components that wrap sub-models and perform any data conversion necessary to support this coupling. 

\begin{figure}
	\begin{center}
	\includegraphics[scale=0.5]{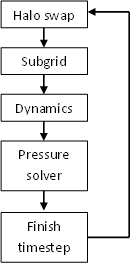}
		\end{center}
	\caption{MONC component group structure}
	\label{fig:monccomponents}
\end{figure}

Like many LES models the simulation proceeds in timesteps, gradually increasing the simulation time on each iteration until it reaches a predefined termination time. The model works on 3D prognostic fields, \emph{u}, \emph{v} and \emph{w} for wind in the \emph{x}, \emph{y} and \emph{z} dimensions, \emph{$\theta$} for the temperature and any number of \emph{q} fields which represent aspects such as moisture and tracers. There are a number of high level groups, each of which contains any number of components, that makes up the structure of a single timestep. Each of these groups must execute sequentially so that one group of components can not start until the previous group has completed. This is illustrated in figure \ref{fig:monccomponents}, where initially all prognostic fields are halo swapped between neighbouring processes and then the sub-grid group of components are called to determine model parameterisations. The dynamics group of components, often referred to as the dynamical core, performs Computational Fluid Dynamics (CFD) in order to solve modified Navier-Stokes equations \cite{lem-science} which is followed by the pressure solver, using either an FFT or iterative approach to solve the Poisson equation. The timestep then concludes with some miscellaneous functionality such as checking for model termination. Microphysics is performed as part of the dynamical core and for every column in the domain the MONC model will call out to CASIM to perform its calculations on that column. 

\begin{figure}
	\includegraphics[scale=0.39]{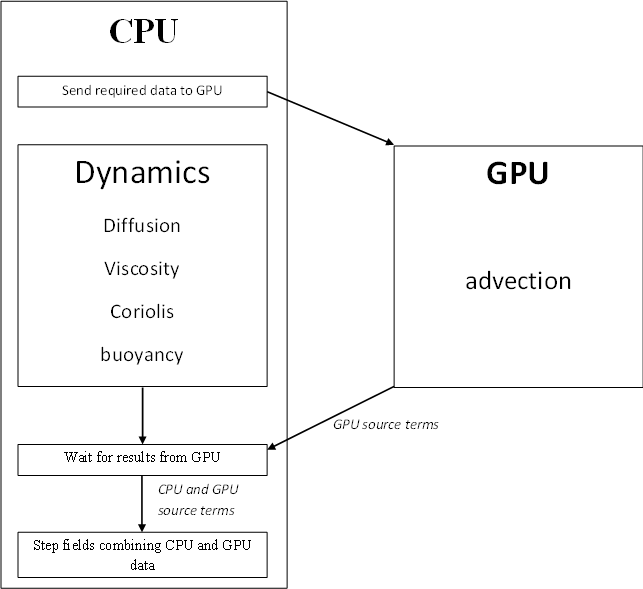}
	\caption{Hybrid dynamics structure}
	\label{fig:dynamics-gpu}
\end{figure}

In \cite{waccpd} we developed an accelerated version of MONC which offloaded the calculations of advection which, previous to integrating CASIM, was the most computationally intensive part of the pure MONC model, onto GPUs. This relied on the fact that the dynamical core proceeds in column fashion, where each component of the dynamical core will process an individual column and contributes the result of its specific calculation by combining (addition operator) it to an initially zero source term for the column of each prognostic field, before going back to the start of the group and calculating for the next column. As we are summing source terms calculated by different components for a column,  this is commutative and associative, which means these dynamical components can be called in any order within the group without impacting correctness. Our theory was that, because we can extract a component and run it on the entirety of a field (all the columns) before, or at the same time as, any other components in the dynamics group without impacting correctness then we could take advantage of this to run MONC heterogeneously on GPUs. To minimise the cost of the copying time to and from the GPU we run both the CPU and GPU concurrently, with them completing separate dynamical core tasks (components) concurrently. Therefore in \cite{waccpd} we split out the advection component from this group and ported it, via OpenACC, to run on GPUs. Figure \ref{fig:dynamics-gpu} illustrates the hybrid execution of the model, where the CPU copied the necessary data to the device asynchronously and then proceeds running the other components in the dynamics group and computing source terms for each column of the fields. At the same time, the GPU receives the required data and then uses this to run the advection kernel and compute advection source terms for the entirety of the field (each column at the same time). Once the CPU has completed the entirety of its own work it then waits for the GPU advection source terms to be made available, combines the CPU and GPU source terms together and then integrates these into the prognostic fields. 

We chose OpenACC for this work because MONC is written in Fortran and-so integration with some other acceleration technologies, such as OpenCL, is more difficult because there is no mature Fortran interface. It was also important, from the Met Office's perspective, to work with a standard and technology that was open, which ruled out CUDA. At the time of this work (2015) implementations of OpenMP 4.0 were less mature and hence OpenACC seemed like an obvious choice. During this work we learnt a lot about using OpenACC and ways to work around issues both with the maturity of the standard and also implementation (Cray compiler), such as deep copying of data which was discussed in \cite{waccpd}. Whilst this work was interesting and proved the general concept of running MONC heterogeneously in this fashion, the amount of computation performed by the advection schemes was just not sufficient to outweigh the cost of data transfer and hence it did not take advantage of the computational power of the GPU. The CPU was often left waiting for the GPU and, in terms of performance, there were no clear benefits to offloading the advection scheme in this manner. However, the CASIM microphysics scheme is far more computationally intensive than advection \cite{slowcasim} and enabling it doubles or even trebles the runtime of MONC. Hence this is a far more interesting candidate for acceleration, using the same heterogeneous approach.

\subsection{Related work}

A number of accelerated versions of climate and weather models, used by the community, have been developed. GALES \cite{gales} is a GPU accelerated Large Eddy Simulation model, implemented in CUDA and working at mixed-precision this was initially based on the CPU only DALES model. With some minor exceptions, the entirety of the computation is performed on the GPU, with the CPU servicing the GPU only and likely sitting idle for much of the run. It is also not clear from \cite{gales} how their model would parallelise over multiple nodes and GPUs which is required in order to tackle the scale of problems that scientists wish to currently investigate.

The Non-hydrostatic Icosahedral Model (NIM) is a weather prediction model developed at NOAA's Earth System Research Laboratory which has been ported to GPUs and the Knights Corner (predecessor to KNL) architectures \cite{nim2014}. The initial port was done using F2C-ACC directives and then the code was updated to also support OpenACC. This project has ported the dynamical core of NIM onto GPUs and KNCs, the paper \cite{nim2014} discusses in detail the optimisations performed to reach the level of performance that they have reached and makes suggestions such as keeping the size of GPU kernels as small as possible. The GPUs (K40s) out performed Knights Corner in their experiments but the performance of the latest generation of Phi, the KNL is known to be significantly better than the KNC \cite{knl-better-knc} so this conclusion may no longer hold true for the latest generation of the hardware.

COSMO is an atmospheric model used for weather prediction by a variety of European organisations and has been ported to run on GPU systems \cite{fuhrer2014towards} as well as experimental ports to KNL \cite{cosmoknl}. Two approaches were adopted for the GPU port, an entire rewrite of the dynamical core and a directive based approach for other parts. The dynamical core, which takes up around 60\% of the overall runtime and is modified infrequency \cite{cosmocug}, was ported to a domain specific language, STELLA \cite{stella}, which separates out the atmospheric model from the architecture specific implementation. Other parts of COSMO, which are less computationally intensive and modified by a wider number of people, have been ported using OpenACC and the community can keep the same code base for these parts without having to learn new languages. The work done in \cite{cosmoknl} discusses how critical it is to performance to get the mapping of MPI processes and threads to the cores correct. They saw a very significant decrease in performance when running at sub-optimal numbers of MPI processes (either too few or too many) which is not necessarily obvious without experimentation. A new project has begun, GridTools, which aims to rewrite STELLA and, amongst other things, enable targetting at KNL \cite{gridtools}. It is still early days but they are hoping to have the one portability layer for numerous architectures to aid in future proofing technology choices.

The Weather Research and Forecasting (WRF) \cite{wrf} model is a numerical weather prediction code developed for atmospheric research. This model has been ported to Knights Ferry (the first Xeon Phi), Knights Corner and now Knights Landing and they are seeing significant performance on the KNL both in comparison to the previous generation KNC and also Broadwell and Haswell CPUs \cite{wrf-knl}. They state that taking advantage of SIMD is an crucial factor for maximising performance and the AVX-512 vector instructions of the KNL are of significant benefit because of the high memory bandwidth keeping the units feed with data, but one needs to ensure that their code can utilise the MCDRAM effectively. 

\section{GPU CASIM}
Whilst previous work done in \cite{waccpd} did not produce the performance benefits we had hoped for, mainly in part due to the limited computation in the advection schemes, the general idea of splitting the dynamics group in this fashion is sound. Offloading specific components and running these concurrently with other components on the CPU, before combining the results together has potential. Since the work of \cite{waccpd}, CASIM has been coupled with MONC and it is at-least five times more computationally intensive than advection. As CASIM is part of the dynamics group we can take advantage of the same general ideas, with the theory being that we will see definite benefit to offloading CASIM because far more computation is being performed than in advection which is likely to amortise the cost of data movement. Due to the existing work and our experience with OpenACC we stayed with this technology to leverage the existing implementation. We also focus on the Cray compiler, not least because the MONC model does not currently compile with the PGI compiler.

\subsection{Loops or the entire code}
\label{sec:entireopenacccode}
The advection code ported to GPUs in \cite{waccpd} was fairly standard in terms of being able to identify a number of computationally intensive loops, decorate these with OpenACC and for these loops to be transformed into our kernels running on the GPU. However with CASIM this was not the case and instead the computationally intensive loops were buried in the code and called many times, for different \emph{q} fields and other values, throughout a column's run. One option was to significantly refactor the code and extract each computationally intensive loop with the host CPU performing other, less intensive, non-floating point operations. In this approach the CPU would perform some work and transfer the required data onto the GPU. Kernels on the GPU would then execute and transfer the required data back to the CPU which would perform some more work and repeat. However this had three major downsides; a significant refactoring of the code required which would very significantly modify CASIM, excessive synchronous data movement onto and off the device for each kernel launch and requiring the CPU to work in lock-step with the GPU instead of taking advantage of the heterogeneous approach developed in \cite{waccpd} and concurrently working on other aspects of the dynamical core. We felt that this approach would involve far too much data movement and the GPU's activity would be very intermittent, instead of keeping it busy.

Instead the decision was made to offload the entirety of CASIM to the GPU, not just the intensive floating point operations but also other loops, conditionals and integer calculations. In this approach all the data movement would be done once en-mass per timestep and then the entirety of the scheme would run on the GPU whilst the CPU performs the rest of the dynamics. Contributing source terms would then be copied back from the GPU to the CPU and integrated. The downside was that a very large amount of code had to be offloaded, including 50 Fortran modules and 123 subroutines, which OpenACC supports but is far less common than the offloading of loops containing floating point calculations only. 

\subsection{OpenACC implementation}
\label{sec:openaccimplementation}
CASIM contains an entry point subroutine, \emph{microphysics\_common} where a specific column (represented by their moisture \emph{q} fields) is provided by the parent model and CASIM operates on this column of data, writing results into source term arguments for each \emph{q} field. The parent model drives the scheme by calling it for each column and we applied OpenACC at this level, around the loops in the two horizontal dimensions.

Listing \ref{lst:openacctoplevel} illustrates the application of OpenACC, where each iteration of the loop is mapped to a single OpenACC vector lane. Calling into \emph{microphysics\_common} this subroutine, along with all the other CASIM subroutines that it calls, are offloaded to the GPU. These subroutines are offloaded via OpenACC 2.0's \emph{routine} directive, with the \emph{seq} clause denoting that there is no further OpenACC parallelism inside this routine (as calculations for a specific column are sequential and tightly coupled.) Effectively this maps each column to a thread on the GPU and all subroutines to be offloaded onto the GPU were marked in this manner. The \emph{async} clause is used with the \emph{parallel} directive here to denote asynchronous launching of the kernel.

\begin{lstlisting}[frame=lines,caption={Top level OpenACC kernel decomposition},label={lst:openacctoplevel}]
subroutine CASIM()
  !$acc parallel async(ACC_QUEUE)
  !$acc loop collapse(2) gang worker vector
  do i = is, ie
    do j = js, je
      call microphysics_common(i,j, ...)
    end do
  end do
  !$acc end loop
  !$acc end parallel
end subroutine CASIM

subroutine microphysics_common(i,j, ...)
  !$acc routine seq
  ...
end subroutine microphysics_common
\end{lstlisting}

Many of CASIM's modules contain a number of global variables which are referenced throughout that module and the entire scheme. In order to handle these the \emph{declare} directive, along with link clause which provides complete control to the programmer over the variable, was used. Whilst explicit directives are required to allocate space for the variable on the GPU and to copy data on and off, we chose this approach for the flexibility that it provides which is most suited to CASIM. Listing \ref{lst:openaccglobalvar} illustrates the \emph{pressure}, \emph{reference\_profile} and \emph{dqv} variables, declared on the GPU via the \emph{declare link} directive. Due to the desire to run GPU kernels asynchronously with the CPU, the \emph{update} directive is used, however this does not allocate memory for variables on the GPU. Therefore data copying is split into two distinct sections, the copying of all variables at initialisation, which copies constants and allocates memory for other non-constants (such as the prognostic fields), and secondly the copying of non-constant input variables to the GPU that vary each timestep (such as \emph{pressure}) and copying back to the CPU result variables (such as \emph{dqv}) for each invocation. The \emph{ACC\_QUEUE} parameter, that we define in code, results in a dependency that data arrival will complete before kernel execution, and only once execution is completed will the copy back of result data (via \emph{update host}) to the CPU begin. The \emph{enter data} and \emph{exit data} directives are required to set the data lifetime to be dynamic, starting a data region and making the variables available to every kernel that executes in this region. At shutdown the \emph{exit data} directive is issued to close the data region and free up memory on the GPU. A total of 250 global variables have been handled in this manner.

\begin{lstlisting}[frame=lines,caption={Handling global variables in OpenACC},label={lst:openaccglobalvar}]
real(wp) :: pressure(:), dqv(:), reference_profile
!$acc declare link(pressure, reference_profile, dqv)

subroutine initialise_micromain()
  ...
  !$acc enter data copyin(reference_profile, pressure, dqv)
end subroutine initialise_micromain

subroutine shipway_microphysics()
  !$acc update async(ACC_QUEUE) device(pressure)
  !$acc parallel async(ACC_QUEUE)
  ...
  !$acc end parallel
  !$acc !$acc update async(ACC_QUEUE) host(dqv)
end subroutine shipway_microphysics

subroutine finalise_micromain()
  !$acc exit data
  ...
end subroutine finalise_micromain
\end{lstlisting}


\subsection{Performance results}
\label{sec:gpuperformance}
Performance tests have been carried out on Piz Daint, both in its previous XC30 configuration (8 core Intel Sandy Bridge CPU and K20X GPU per node) and current XC50 configuration (12 core Intel Haswell CPU and P100 GPU per node.) A standard MONC test case for modelling stratus cloud has been used and in this section we consider both the simplest CASIM warm configuration which involves computation with 5 moisture \emph{q} fields and also the more complex cold case which requires 18 moisture \emph{q} fields and involves significantly more computation than its warm counterpart. They are interesting to compare because the cold case is far more computationally intensive, so might make better use of the GPU, but also involves copying far more data (due to the extra number of fields) to and from the GPU which will have an overhead. We vary the number of horizontal columns in two dimensions (hence the number of GPU threads), fixing the vertical column height to be 60 grid points. The Cray compiler was used for all tests and the presented results were averaged over three runs.

\begin{figure}[!htbp]
	\includegraphics[scale=0.38]{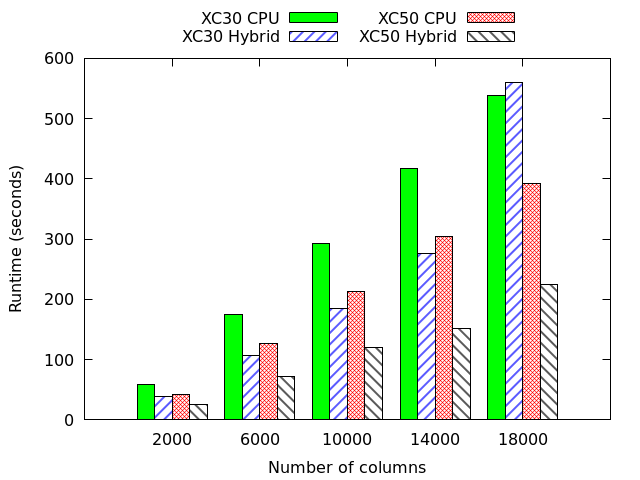}
	\caption{XC30 vs XC50 runtimes for warm stratus test case on MONC with CASIM}
	\label{fig:xc30vsxc50}
\end{figure}

Figure \ref{fig:xc30vsxc50} illustrates the model runtime for MONC, running CASIM, both in CPU only and hybrid GPU mode on both the XC30 and XC50 configurations of Piz Daint utilising one CPU core on a single node. From the homogeneous perspective it can be seen that there is a significant increase in CPU performance going from Sandy Bridge to Haswell but there are also a number of other observations that can be made from graph in terms of heterogeneous performance. On the XC30 it can be seen that the hybrid code exhibits a very large increase in runtime between 14000 and 18000 vertical columns, actually taking longer on 18000 columns than the CPU version. Fundamentally this was due to a lack of registers, the default setting was to allocate 128 registers per thread giving a theoretical occupancy on the K20X of 0.25. We reduced the register count to 64 registers per thread which then improved the theoretical occupancy to 0.5. The K20X has 14 Streaming Multiprocessors (SMs) \cite{p100-specs}, each capable of executing up to 2048 threads, but with a theoretical occupancy of 0.5 this becomes 1024, and 1024 * 14 = 14336 concurrent threads maximum. As we are mapping each vertical column to a GPU thread, at exactly 14337 columns we see a large jump in runtime on the K20X because columns now have to be split into two sequential batches rather than running concurrently. We don't see this issue with the P100 because, whilst the number of registers in each SM and the maximum number of threads an SM can execute is the same, the P100 is equipped with 56 SMs \cite{p100-specs} rather than 14 in the K20X. Therefore the P100 affords us a limit of 57334 threads and hence that is the maximum theoretical number of concurrently executing columns possible. Due to these extra registers we were able to increase the number of registers per thread to 128 which resulted in a performance increase in comparison to CASIM running on the K20X. For instance running in hybrid mode with 14000 vertical columns there was a reduction in MONC runtime of 65\% on the XC30, whilst on the XC50 the runtime halves in comparison to its homogeneous counterpart.

\begin{figure}
	\includegraphics[scale=0.38]{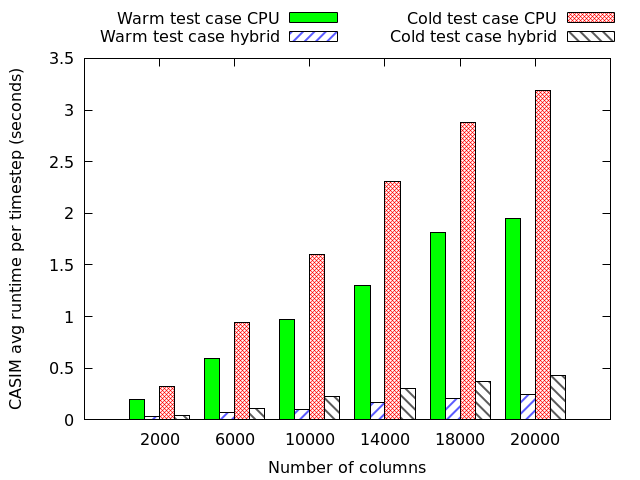}
	\caption{Average CASIM runtime per timestep using on one CPU core}
	\label{fig:warmvscoldgpu}
\end{figure}

However the overall picture illustrated by figure \ref{fig:xc30vsxc50} only tells half the story. Whilst the overall runtime speed-ups look impressive for the heterogeneous MONC model running CASIM on the GPU, we effectively have a two way concurrency because CASIM is running on the GPU at the same time that the MONC model is doing the rest of its advection on the CPU. It is therefore important to look further to answer the question of whether the GPU is actually giving a benefit here in contrast to using the same approach but running the scheme concurrently on another CPU core instead. We looked more closely at the performance of CASIM itself to understand whether running it on the GPU provides any performance benefit, or if the drop in runtime is mainly because of the dynamics concurrency. Figure \ref{fig:warmvscoldgpu} illustrates the average time taken by CASIM per timestep for different numbers of vertical columns with both the warm and cold stratus test cases on the XC50 against one CPU core. It can be seen from this graph that running CASIM on the GPU is significantly faster than running on the CPU, for instance executing the cold stratus test case over 20000 vertical columns on the GPU is 7.4 times faster than running it on one CPU core only. Even though the warm test case is less computationally intensive, a similar speed up is achieved in comparison to it also running on a single CPU core. 

\begin{figure}
	\includegraphics[scale=0.38]{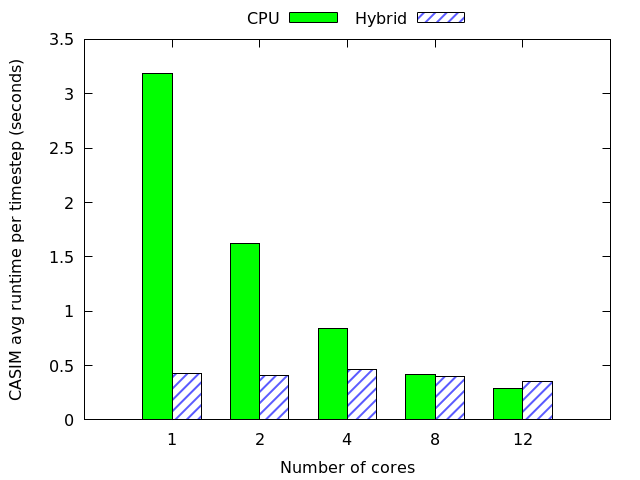}
	\caption{Average CASIM runtime per timestep for cold test case over 20000 vertical columns as the number of CPU cores is modified}
	\label{fig:gpucores}
\end{figure}

The results in this section have so far concentrated on running MONC, and CASIM, on a single Haswell core. However the Haswell processors of Piz Daint are equipped with twelve CPU cores per package and as such it is not necessarily realistic to just compare against runs involving one core only, with the rest remaining unused. Figure \ref{fig:gpucores} illustrates the average CASIM runtime per timestep over 20000 vertical columns for the cold stratus test case as we modify the number of CPU cores in use (weak scaling), effectively decomposing the global domain over these cores each running MONC concurrently. In the case of the hybrid version of our experiments, where we run MONC on the CPU and CASIM on the GPU, each core shares the same GPU via MPS's multiplexing, effectively running multiple kernels on the GPU concurrently. From figure \ref{fig:gpucores} it can be seen that the very significant speed advantage to running CASIM on a GPU in comparison to one CPU core doesn't hold as we increase the number of cores. The break even point is around 8 CPU cores and at 12 CPU cores it is actually faster to run CASIM on the CPU instead of the GPU. Columns in CASIM are independent, hence the code is embarrassingly parallel and scales very well (there is an 11 times speed up over 12 cores in comparison to 1 core), and so we are utilising more of the resources of the CPU whilst the GPU is being utilised exactly the same, just driven by different CPUs. Whilst the runtime on the GPU is fairly flat, in comparison to being driven by one CPU core the average time per timestep does decrease slightly over twelve cores to 81\% of the runtime driven by one core.

Whilst the results in figure \ref{fig:gpucores} do argue that running CASIM over the entirety of the CPU cores, rather than the GPU, is advantageous, it should be noted that our approach to offloading is a hybrid one where the parent model will run concurrently with CASIM. Due to this concurrency the overall runtime of MONC is still less over twelve CPU cores when run hybrid in comparison to running the entirety of MONC and CASIM sequentially on the CPU. To understand the reasons for the performance behaviour of the GPU version, table \ref{fig:timeindata} illustrates the relative time spent by the GPU copying data on, off and doing computation. It can be seen that even for larger numbers of columns the vast majority of the runtime is being spent in the kernel and data transfer actually only takes up a small fraction of the overall time. With the warm test case, which involves transferring considerably fewer fields but also less computation, interesting the data transfer times are not too dissimilar from the cold test case and the time spent in execution by the kernel is slightly lower as an overall percentage than the cold test case. However both these configurations are spending the vast majority of their runtime during execution on the GPU rather than data transfer. One can therefore conclude that there is sufficient computation to amortise the cost of data transfer, unlike the previous work done in \cite{waccpd}. 

\begin{center}
\begin{figure}
	\begin{tabular}{ | c | c | c | c | c | }
		\hline
		Column size & Config & To GPU & Kernel & From GPU \\ \hline
		2000 & Warm & \specialcell{1.5ms \\ 5\%} & \specialcell{29ms \\ 93\%} & \specialcell{0.8ms \\ 3\%}\\
		2000 & Cold & \specialcell{1.82ms \\ 4\%} & \specialcell{39ms \\ 94\%} & \specialcell{0.74ms \\ 4\%}\\ 
		10000 & Warm & \specialcell{7ms \\ 7\%} & \specialcell{88ms \\ 88\%} & \specialcell{4.6ms \\ 5\%} \\ 
		10000 & Cold & \specialcell{11.2ms \\ 5\%} & \specialcell{214ms \\ 93\%} & \specialcell{4.6ms \\ 2\%}\\ 
		20000 & Warm & \specialcell{18ms \\ 7\%} & \specialcell{224ms \\ 90\%} & \specialcell{8ms \\ 3\%} \\
		20000 & Cold & \specialcell{22.56ms \\ 5\%} & \specialcell{395ms \\ 93\%} & \specialcell{8.1ms \\ 2\%}\\
		\hline		
	\end{tabular}
	\caption{Time spent in different phases of kernel launch}
	\label{fig:timeindata}
\end{figure}
\end{center}

\begin{figure}
	\begin{center}
	\includegraphics[scale=0.38]{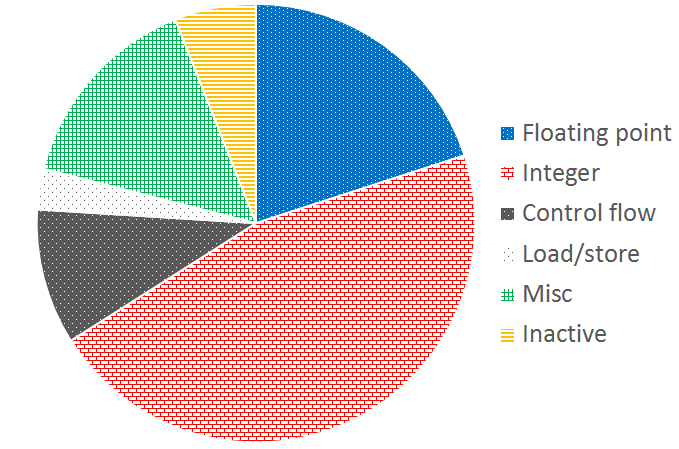}
	\end{center}
	\caption{Distribution of instruction type during typical CASIM kernel execution}
	\label{fig:gpuinstructions}
\end{figure}

Figure \ref{fig:gpuinstructions} illustrates the distribution of instruction types during execution of a CASIM kernel. At 46\% of the overall instructions, integer arithmetic is the far most significant type of instruction executed by CASIM. Floating point operations, what a GPU is ideal for, is the second most common form of instruction but at only 20\%. The kernel is only idle, due to warp divergence or conditional prediction, for 6\% of its time. Initially we had thought that warp divergence might be a significant limitation upon CASIM's GPU performance, but this is only a fraction of the overall instruction distribution. Far more significant is that almost half of the instructions issued to the GPU are integer instructions rather than the floating point operations that GPUs are so suited towards.

A major limitation of offloading the entirety of CASIM which we did not anticipate is that of memory limits. Currently 20000 vertical columns (1.2 million grid points) is the maximum data size possible and beyond this the GPU runs out of memory. But even for the largest run, there is only around 256 MB of data copied onto the GPU as an input for CASIM on each timestep. However upon initialisation a large number of data structures are allocated which store temporary data, for instance the different distributions of moisture, per column. CASIM has been written in such a way that many temporary variables are used in processing a single column, representing different values for different states up and down the column. On the GPU each column is being run concurrently by separate threads, hence we have to replicate all these temporaries for every column so that each thread has its own effectively private local data areas. This replication, which is unavoidable in our approach, results in a significant memory overhead and hence we hit the memory limit of the GPU.

\subsection{Optimisations}
\begin{figure}
	\includegraphics[scale=0.38]{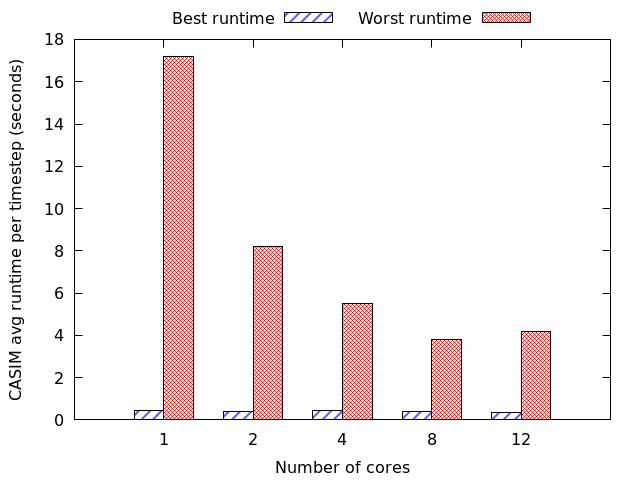}
	\caption{Average CASIM runtime per timestep variance from best to worst configuration over 20000 vertical columns}
	\label{fig:gpuvariance}
\end{figure}

The results presented in section \ref{sec:gpuperformance} represent the best possible configuration of CASIM, ported to GPUs with OpenACC. As discussed in section \ref{sec:openaccimplementation} we optimised aspects such as the data transfer to limit the data that must be copied on and off the GPU which is fairly standard and based upon lessons we had learnt in \cite{waccpd}. We also found empirically that 128 registers per thread gave the best performance on the P100. However far more important for performance was choosing an appropriate number of thread blocks (the number of gangs in OpenACC) and selecting an appropriate number of threads per block (vector length in OpenACC.) These are set by optional clauses in the \emph{parallel} OpenACC directive and it was found that omitting this and relying on default values resulted in very poor performance. Additionally the optimal configuration varied significantly based on the number of vertical columns and number of cores sharing the GPU. This variation is illustrated in figure \ref{fig:gpuvariance}, showing the best performance on the GPU per core and the worse performance on the GPU per core over 20000 vertical columns for the cold test-case. The difference is significant and the only way of deducing correct settings was by experimentation.

We ran over 8000 permutations of different configurations and generally speaking at smaller numbers of vertical columns (threads) then more, smaller, thread blocks are desirable. As one increases the number of columns (and hence threads) then fewer larger thread blocks are instead optimal. With smaller amounts of data to compute per column (i.e. lower values for the height of columns or the warm test case instead of the cold test case) then configurations utilising fewer thread blocks performed best. As one increases the number of cores, which effectively lowers the number of threads per kernel, then again the configuration of fewer, larger thread blocks is preferable.

\subsection{Programming challenges}
For the work described in this section the Cray compiler, environment version 8.5.5 was used which is the standard compiler used by MONC and CASIM on homogeneous Cray systems. Whilst this compiler implements the OpenACC 2.0 standard fully we found a number of bugs and limitations with the compiler that had to be worked around. This is not hugely surprising because, by offloading very many entire subroutines, we were using OpenACC in a much more complex manner than many other codes which simply offload computational loops. The short comings and associated workarounds that we describe in this section have not been documented elsewhere and most commonly were when the Cray compiler generated incorrect PTX instructions, often at higher levels of optimisation. 

\subsubsection{Passing arrays of derived types}
When a number of CASIM routines where compiled, the assembler reported the error \emph{Arguments mismatch for instruction 'mov' Unknown symbol 't\$5' Label expected for forward reference of 't\$5'}. This was when it was attempting to assemble the PTX code. These routines all accept an array of a derived type as a dummy argument and the compilation of these arguments is erroneous as other arrays of non-derived types compile without issue. Inspection of the PTX code revealed that when an array must be passed as an argument to an accelerator routine, a data structure is used which acts as an array descriptor storing information such as the array’s bounds, number of elements, pointer to the data, and other fields. When an array of derived type is passed then broken PTX code is generated by the Cray compiler for the preparation of this data structure. 

The problem was solved by wrapping the offending array into a derived type itself, where the original array is now a member of the wrapper type and only a single instance of this wrapper type is passed to the routine. We found it also necessary to store the length of the array in this wrapper type because this approach was required for a number of different routines. Each of these routines originally accepted arrays of different lengths and because the use of allocatable members in derived types code is discouraged in OpenACC, explicitly storing the length was instead adopted.

\subsubsection{Large arguments}
As has been described in section \ref{sec:entireopenacccode} the entirety of the scheme has been offloaded to the accelerator and the main loop over vertical columns was wrapped in an OpenACC accelerator region. Because the compiler used in this project employs NVIDIA CUDA as the OpenACC back-end, the accelerator region is implemented as a CUDA kernel function which is subsequently executed by the GPU. The variables referenced in the lexical scope of the accelerator region, i.e. in its immediate body and not in the procedures it calls, are passed through as arguments to the CUDA kernel function implementing the region. However, when a certain threshold size or number of arguments is exceeded the compiler passes these variables differently by packing them into a contiguous buffer in memory on the accelerator and passing the pointer. Code on the GPU subsequently unpacks this buffer which is transparent to the user. This feature is referred to as \emph{Large arguments} and the mechanism requires support from both the host side and the accelerator side. The host side of this feature is not implemented by the Cray compiler and generates the runtime error \emph{ACC: craylibs/libcrayacc/acc\_hw\_nvidia.c:915 CRAY\_ACC\_ERROR - Large args not supported}. Neither the nature of the threshold, i.e. whether it applies to the overall number or the size of arguments, nor any specific numerical value were mentioned in documentation. Empirically it was found that the feature is triggered when the CUDA kernel has more than 532 arguments.

Because CASIM uses enough variables in the lexical scope of the accelerator region to trigger this error it became necessary to reduce the number of CUDA kernel function arguments used by our code. Two factors were found to influence the number of kernel arguments most significantly:
\begin{enumerate}
\item the number of distinct module arrays used in the lexical scope of the accelerator region
\item the number of times each such array is referenced in the lexical scope of the accelerator region
\end{enumerate}
The total count is impacted most significantly due to arrays because they often utilise multiple arguments. In the simple warm stratus test case, there are 26 distinct 3 dimensional fields (representing the original 5 \emph{q} fields) and 34 3-dimensional aerosol fields passed into CASIM's accelerator region. Each field uses up to ten arguments and hence these alone occupy several hundred CUDA kernel arguments. To reduce the total count we packed these fields into a four dimensional array and then referencing them on the GPU directly from this structure, which brought the argument count below the \emph{large arguments} threshold and avoided the error. The fact that large arguments is not implemented by the Cray compiler suggests not many, if any, codes push Cray's OpenACC implementation to its limits in this aspect.

\subsubsection{Assigning variables in conditionals}
In certain subroutines local variables were only assigned and used in conditionals, such as in listing \ref{lst:openaccvarcond} where variable \emph{dm\_3} is local to some subroutine but no value is assigned to it by default. If a specific conditional is met then a value is assigned and later in the code if the same logical conditional is met (i.e. the variable will never be read unassigned) then it is used. At any optimisation level this resulted in the error \emph{INTERNAL COMPILER ERROR: ptx\_cg::generate\_code\_for\_sym(): use before def of global temp}. The work around was to ensure that local variables are always assigned a value (often zero) at the start of the specific subroutine. This is of course good practice anyway, but worth noting when working on an existing code base to determine whether there are any variables of this form.

\begin{lstlisting}[frame=lines,caption={Assigning and using a variable in conditional only error},label={lst:openaccvarcond}]
...
if (some_condition) dm_3=5.435
...
if (some_condition) othervariable=dm_3
\end{lstlisting}

\subsubsection{Profiling}
NVDIA's nvprof was used to profile the kernels on the GPU, however a number of key limitations were identified in this tool. Firstly, due to the Cray compiler converting the entirety of the code to a single CUDA kernel, it was not possible to drill down and understand the behaviour of the OpenACC code for each individual routine being executed. This would have been very useful in our work but because most OpenACC codes are based around fairly simple computationally intensive loops then it is likely not been prioritised. 

The other major limitation was in collecting metrics and events which are required for in-depth analysis of the kernel. When collecting either of these with local problem sizes larger than 40 by 50 columns then the code failed with an out of memory error. With these smaller model runs, collecting the information also very dramatically increased the execution time on the GPU, so it is not entirely clear if the information obtained was representative of a real run or not. It is believed that most often CUDA kernels are far simpler and smaller than the one in use here, and the large size of our kernel is responsible for this shortcoming. It is entirely possible that the developers of nvprof simply have not designed their tool a single kernel of such size. Once again, if it was possible to split the OpenACC code up into multiple kernels, then that would most likely assist in this situation too.

\section{Knights Landing CASIM}
Programming for Knights Landing (KNL) is, in many ways, more straightforward as this technology supports direct execution of MPI codes and/or CPU threads using technologies such as OpenMP. A version of CASIM using OpenMP for threading was implemented and similar to the OpenACC version orients the parallelism around each column iteration, so that processing columns can be done independently by threads. Module global variables, mentioned in section \ref{sec:openaccimplementation}, were handled via the \emph{threadprivate} directive, placed next to the variable definition. The \emph{copyin} clause was provided to the \emph{parallel} directive, which both allocates space for threads copies of allocatable arrays and also copy in any prerequisite values from the \emph{master} variables into the private copy for each thread. The SIMD OpenMP directive was applied to computationally intensive loops in the code that work up or down the column.
\subsection{Performance}
\label{sec:knlperformance}
\begin{figure}
	\includegraphics[scale=0.38]{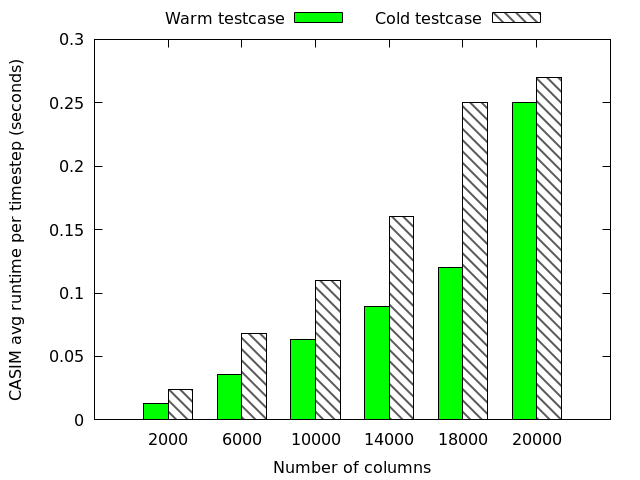}
	\caption{Average CASIM runtime per timestep on the KNL}
	\label{fig:knlgeneral}
\end{figure}

MONC and CASIM, with the same warm and cold stratus test cases as described in section \ref{sec:gpuperformance} were executed on the 64 core model 7210 KNL, which are part of ARCHER the UK national supercomputer (the KNL aspect of this machine is a Cray XC40.) Each node is equipped with 96GB of main memory and 16GB of on chip MCDRAM which was configured to use as a cache for accessing the main memory and run in quadrant mode. Figure \ref{fig:knlgeneral} illustrates the performance of CASIM when running both cold and warm stratus test cases on the KNL and the presented results were averaged over three runs. With the warm test case it can be seen that there is a large jump in runtime between 18000 vertical columns and 20000 vertical columns but this configuration signified another important aspect too. Namely that up to and including 18000 columns running with 256 processes, 4 MPI processes per core (with hyper-threading) resulted in the best performance. However from 20000 columns, still using 4 way hyper threading, where each physical core runs one MPI process and each process executes four threads was optimal. 

\begin{figure}
	\includegraphics[scale=0.34]{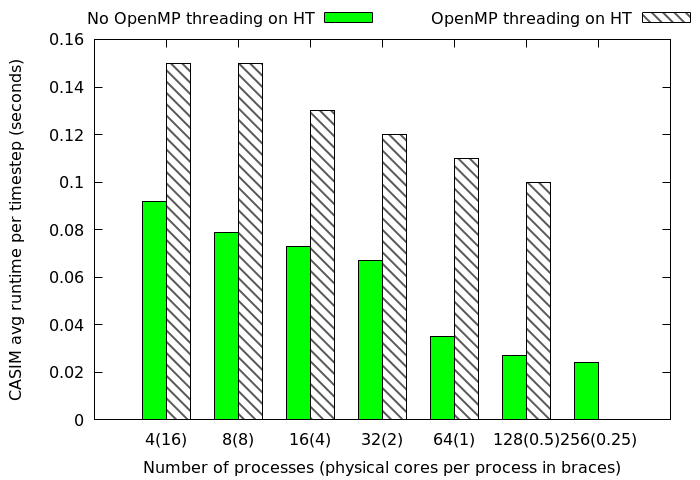}
	\caption{Average CASIM runtime per timestep on the KNL with the cold test-case over 2000 columns}
	\label{fig:knlsimple}
\end{figure}

This same hyper-threading behaviour was also true for the cold test case and this point occurred sooner, at 12000 vertical columns. Figure \ref{fig:knlsimple} illustrates the performance of a CASIM timestep for the cold test-case with 2000 columns, it can be seen that enabling hyper-threading and running OpenMP threads on these results in a runtime increase of approximately three times and there is a performance penalty when we run less than 64 processes and thread across the cores instead. The results of 128 and 256 processes denote where we have enabled hyper-threading but run MPI processes on these hyper-threads instead of OpenMP threads which was found to give optimal performance for this problem and domain size. Similarly to the warm test-case, for 20000 columns enabling hyper-threading and running threads across these (in contrast to processes which is optimal at 2000 columns) was the best strategy. For all these runs the OpenMP \emph{guided} scheduling was found, by experimentation, to be best scheduler. The \emph{dynamic} scheduler was a close second choice but the default, \emph{static}, scheduling produced poorest performance and effectively negated the benefit in enabling hyper-threading for larger column counts. There are a significant number of conditionals within the processing of each column, and it was found that the amount of computation associated with each column is therefore not equal. Hence a static scheduling strategy will result in significant work imbalance which is why the performance it affords is so poor in contrast to a dynamic approach.

\begin{figure}
	\includegraphics[scale=0.38]{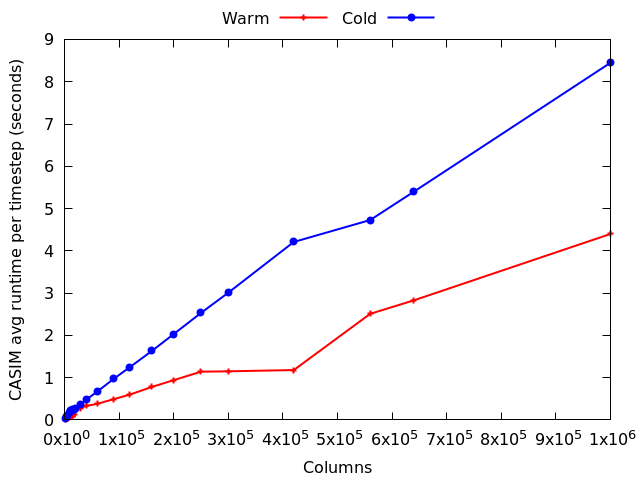}
	\caption{Average CASIM runtime per timestep on KNL with large data sizes}
	\label{fig:knllarge}
\end{figure}

The column numbers described in figure \ref{fig:knlgeneral} and that we have been concentrating on so far represent very large local domain sizes per node for traditional MONC runs. With the column height of 60, the smallest number of vertical columns (2000) presented would result in 120000 grid points, whereas the largest system would result in 1.2 million grid points. In \cite{easc} we ran with 65536 local grid points per CPU core which is typical of current users of the code and advised best practice. However using the KNL it is possible to scale up to far larger data sizes and this is potentially where we might see additional benefits to using the architecture. Figure \ref{fig:knllarge} illustrates average CASIM runtimes per timestep as we scale up the problem size by increasing the number of vertical columns, distributed amongst the 64 KNL cores running 4 hyper-threads per core. The maximum size is a million columns (60 million grid points), beyond which the node runs out of memory as both the MCDRAM (running as a cache) and node main memory has been exhausted. This memory limit is more a facet of the parent MONC model, which must store additional prognostic fields (such as wind and temperature) which CASIM does not use, and also maintains three versions for each prognostic field. With smaller data sizes as we scale up the problem size the runtime increases reasonably proportional, for instance when we go from 2000 vertical columns (120000 grid points) to 4000 columns (240000 grid points) the runtime doubles. However when we run with 60 million grid points, increasing the problem size 500 times, the runtime has only increased 237 times in comparison. Therefore as we scale up the problem size the runtime increase becomes less. 

\begin{figure}
	\includegraphics[scale=0.38]{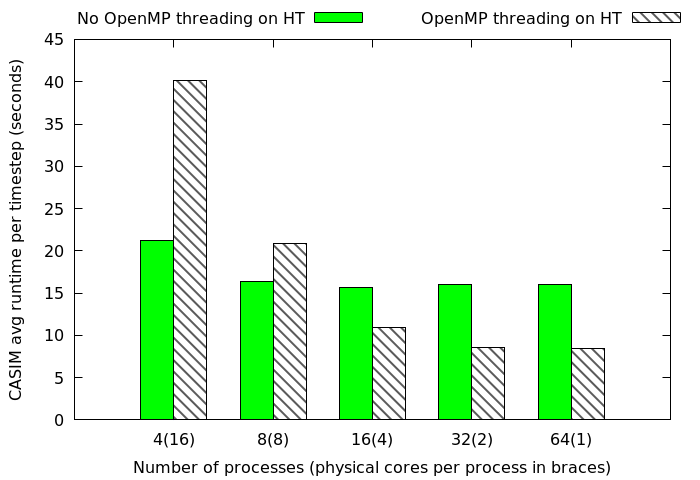}
	\caption{Average CASIM runtime per timestep on KNL with the cold test-case over 100000 columns}
	\label{fig:knllarge_tp}
\end{figure}

Figure \ref{fig:knllarge_tp} illustrates the impact of different process-thread configurations for the largest run of 100000 columns (60 million grid points.) It can be seen that either running with 64 processes, each utilising 4 threads on the hyper-threads or 32 processes, each utilising 8 threads on the hyper-threads is the optimal configuration. As we reduce the number of processes beyond this with hyper-threading enabled then the performance decreases. Interestingly with hyper-threading disabled, whilst the best possible performance is poorer than utilising hyper-threading, the impact of reducing the number of process is far less severe. It was not possible to do tests with 128 or 256 processes on hyper-threads as an Out Of Memory (OOM) error was reported, this is not hugely surprising as this largest configuration is running right at the memory limit and more processes, in contrast to threads, will result in additional memory overhead.

Table \ref{tbl:memoryconfiguration} illustrates the average CASIM runtime per timestep (cold test-case over 10000 columns) when MCDRAM was used in different configurations. The \emph{Flat mode numactl} entry illustrates the case where MCDRAM is used in flat mode but all of the user's memory is explicitly placed into MCDRAM. It can be seen that taking advantage of MCDRAM provides a significant performance benefit to not utilising it (flat mode), and the performance one gains from cache mode is comparable to explicitly placing all memory in MCDRAM. From this experiment we therefore conclude that for CASIM, utilising MCDRAM in cache mode is optimal and this matches generally accepted wisdom \cite{knl-book}.  

\begin{table}[h]
	\centering
	\begin{tabular}{ | c | c | }
		\hline
		Configuration \quad&\quad Average CASIM runtime per timestep (s) \quad  \\
		\hline			
		Flat mode \quad&\quad 0.27\\
		Flat mode numactl \quad&\quad 0.11\\	
		Cache mode \quad&\quad 0.11\\
		\hline
	\end{tabular}
	\caption{MCDRAM configuration average CASIM runtime per timestep for cold test-case over 10000 columns}
	\label{tbl:memoryconfiguration}
\end{table}

\section{Comparison with CPU CASIM}
\begin{figure}
	\includegraphics[scale=0.38]{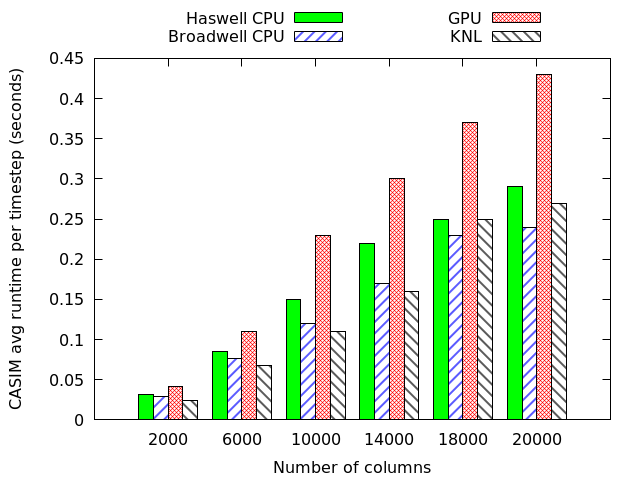}
	\caption{Average CASIM runtime per timestep cold test case comparison between CPUs, GPUs and KNLs}
	\label{fig:allcomparison}
\end{figure}

In addition to the runs conducted in sections \ref{sec:gpuperformance} and \ref{sec:knlperformance}, we have also performed CPU only runs for comparison using a Haswell (12 core) processor and Broadwell (18 core) CPU. Haswell tests were performed by utilising the CPU only of Piz Daint, an XC50, and the Broadwell machine used is the UK Met Office's XC40. One of the key questions driven by this work is what is the best choice for models like CASIM, should we invest in accelerators such as GPUs and KNLs, or instead focus more on powerful homogeneous CPU only machines.Figure \ref{fig:allcomparison} compares the average runtime of a single CASIM timestep using the entirety of a Haswell CPU (all 12 cores), a full Broadwell CPU (all 18 cores), a P100 GPU and KNL-7210. From these results it can be seen that the 18 core Broadwell and the KNL versions perform the best, the Haswell CPU is in third place but breaks even with the KNL at 18000 columns. It can be seen for all configurations that the GPU version performance is the poorest.

\begin{figure}
	\includegraphics[scale=0.38]{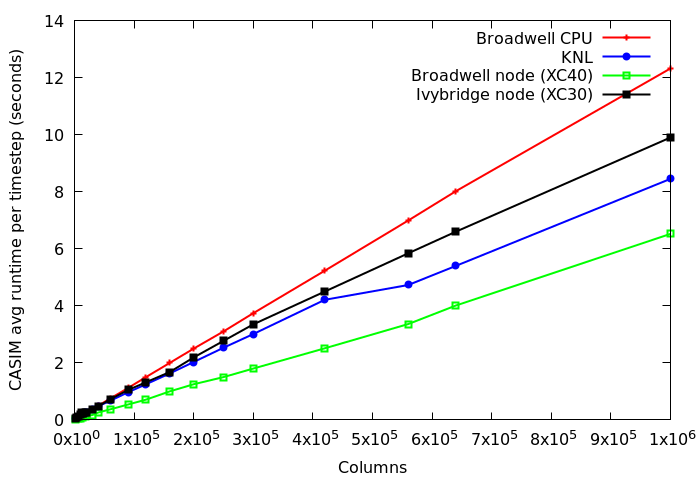}
	\caption{Average CASIM runtime per timestep cold test case comparison between Broadwell, Ivy bridge and KNL on larger column counts}
	\label{fig:broadwellknllong}
\end{figure}

It was noted in section \ref{sec:knlperformance} that the KNL seems to favour working on larger domains, such as 60 million grid points, in comparison to smaller amounts of data. Figure \ref{fig:broadwellknllong} illustrates a comparison between CASIM running on the 18 core Broadwell CPU and the KNL as we scale the number of vertical columns and hence the number of grid points. Whilst the Broadwell CPU and KNL exhibit similar performance on smaller numbers of columns the break even point is 30000 vertical columns and after this point the KNL performance is obviously beating that of the Broadwell. As we reach one million columns (60 million grid points) there is a significant difference and the runtime per CASIM timestep on the KNL is 65\% of that on the Broadwell CPU. It is believed that this behaviour is due to the MCDRAM and utilising this as a cache is advantageous in comparison to the CPU for larger data sizes. A comparison between sockets and the KNL is not necessarily fair as XC CPU machines have two CPUs per node. In figure \ref{fig:broadwellknllong} we also compare an XC40 node (two Broadwell 18 core CPUs) and an XC30 node (two 12 core Ivy Bridge CPUs.) It can be seen that the XC40 node performs the best for all configurations, this is not hugely surprising as CASIM is embarrassingly parallel and hence scales very well and we are contrasting 36, more powerful, Broadwell cores against 64, simpler, KNL cores. CASIM on the KNL outperforms the Ivy Bridge XC30 nodes (ARCHER) as this has fewer cores and is of an older generation than the Broadwell CPUs. From these results we therefore believe an important message for potential users of CASIM on the KNL is that loading it up with a significant amount of computation (and hence number of grid points) is the best approach.

A major limitation of the OpenACC GPU version of CASIM was the significant amount of memory required to hold temporary variables, a distinct copy required per thread. However because on the KNL we are only running with a small number of threads in comparison to the P100 then the memory overhead is far less. This is due to a fundamental difference in the architectures, but one that greatly impacts the grid size scalability of the code.

\section{Conclusions and further work}
In this paper we have explored the applicability of both GPUs and KNLs to the microphysics model CASIM. Due to the inherent design of the code our approach taken for porting onto GPUs was unusual where we offloaded not only the computationally intensive aspects but also other aspects too such as conditionals, loops and integer arithmetic. Whilst there was some benefit to doing this in comparison to a single CPU core, the fact that the GPU was not continuously performing floating point calculations and the significant memory requirements meant that there were some inherent shortcomings with this approach. From the findings of this work we believe that OpenACC and its implementations are mature enough to use for offloading significant code to the GPU, however one needs to ensure that the GPU will not spend a significant fraction of its time working on non-floating point operations and also that any additional memory requirements are well understood. It is our feeling that if we could significantly increase the domain size on the GPU, to a similar size that we have run on the KNL, then it would be more advantageous but this is not possible due to the replication of temporary variables hitting the memory limit.

In comparison it was far easier to take advantage of KNL and the performance obtained was more favourable in comparison to the latest generation CPUs, although a Broadwell XC40 node outperforms the KNL. We found that the key here was to load as many grid points and hence computation onto the KNL, in which case it significantly outperforms a single latest generation CPU in a socket to socket comparison. Irrespective best performance was obtained when we ran at least one MPI process per core, at smaller domain sizes enabling 4-way hyper-threading and running a process per hyper-thread was optimal and at medium to larger numbers of grid points instead placing one process per physical core and threading over the hyper-threads gave best performance. We therefore conclude that KNLs and Cray machines with this technology are more suited to accelerating CASIM, not least because they are more general purpose than GPUs and hence can handle the other, non computationally intensive floating point aspects of the code.

When discussing the performance on the KNL in section \ref{sec:knlperformance} we focused just on CASIM rather than quoting any overall runtime numbers for MONC with CASIM. The reason for that is that the rest of the MONC model has not been optimised for CASIM and as such does not utilise the architecture to its full potential. Important future work will be, based on what we have learnt here, further optimisation of other MONC components, such as the advection schemes, for the KNL. Additionally whilst we have applied SIMD directives to the computationally intensive loops in the KNL code for a vertical column, we have not explicitly aligned the data. As per \cite{knl-book} it is preferable to ensure arrays especially are aligned to 64 byte boundaries and as such further development of the code to investigate and ensure this is desirable. 
\section*{Acknowledgements}
The authors would like to express their gratitude to the Swiss National Supercomputing Centre, CSCS, who have kindly given us access to and time on Piz Daint for this work. This work was funded under the embedded CSE programme of the ARCHER UK National Supercomputing Service (http://www.archer.ac.uk)

\end{document}